\documentclass{iopart}
\usepackage[dvips]{graphicx}
\begin{document}

\title[Electromagnetic signature of jets]{The electromagnetic signature of jets}

\author{Charles Gale\dag, T. C. Awes\S, Rainer J. Fries\P and Dinesh K. Srivastava\ddag}

\address{\dag\ Department of Physics, McGill University, 3600 University Street,
Montreal, QC, H3A 2T8, Canada}

\address{\S\ Oak Ridge National Laboratory, Oak Ridge, Tennessee 37831, USA}

\address{\P\ School of Physics \& Astronomy, University of Minnesota, 16 Church
Street S.E., Minneapolis, Minnesota 55455, USA}

\address{\ddag\ Variable Energy Cyclotron Centre, 1/AF Bidhan Nagar, Kolkata 700
064, India}

\begin{abstract}

Recent RHIC data have suggested an interesting scenario where jets, after
being formed in the very first instants of the nuclear collision, interact
strongly and are
absorbed by the hot and dense matter subsequently created. 
In this respect, electromagnetic signals constitute another class of
penetrating radiation.  We first propose the study of large mass dileptons induced 
by the passage of high-energy
quark jets through quark-gluon plasma. We find that the yield due to the jet-plasma
interaction gets progressively larger as the collision energy
increases: it is slightly less than the Drell-Yan contribution at RHIC energies, and
it is a factor of 10 larger
than Drell-Yan at the LHC. In a related investigation, 
we then propose to study dilepton-tagged jets. 
We estimate quantitatively the various
background sources, and identify windows where those are sub-dominant.

\end{abstract}




\section{Introduction}
Relativistic heavy ion collisions currently constitute a
paradigm for studies of matter under extreme conditions. In a complex many-body
environment whose existence spans a wide range of dynamical
variables, several complementary observables are needed to identify new physics.
Hard hadronic probes, as measured by the current generation of RHIC experiments,
seem to support a very interesting scenario where the QCD jets being formed in
the first instants of the collision are being quenched by the hot and dense
matter \cite{jet}. Albeit compelling, the experimental evidence still needs to
be interpreted in a quantitative and systematic theory in order to clearly
distinguish effects in the final state from effects in the initial state, for
example \cite{cgc}. Jet-quenching is in fact a strong jet-plasma interaction.
If such phenomena occur, the same interactions should yield other products.
Recalling the need to identify new phenomena with different measurements, we
have calculated the production of electromagnetic radiation resulting from the
passage of jets through a quark-gluon plasma. An observation of this new source
would confirm the occurrence of the conditions suitable for jet-quenching to take
place.

\section{Formulation and results}
We propose that the phase space distribution of partons in the hot and dense
medium can be decomposed as
\begin{eqnarray}
f (\mbox{\boldmath $p$}) = f_{\rm th} (\mbox{\boldmath $p$}) + f_{\rm jet}
(\mbox{\boldmath $p$})\ ,
\end{eqnarray}
where the plasma component is approximated by a thermal distribution with a
temperature $T$: $ f_{\rm th} (\mbox{\boldmath $p$}) = \exp (-E/T)$. The jet
component is calculated from perturbative QCD, and is limited to a region
$p_{\rm T} \gg$ 1 GeV. This separation is tenable kinematically, as jet spectra
fall of as a power law and can thus be differentiated from the thermal
component. The phase space distribution for jets can be obtained from
perturbative QCD calculations, with an appropriate normalization  allowing for
the space-time extent of the emitting source \cite{lg}. Given the parton
distribution functions, one may calculate the production rate for real photons
from jet-plasma interactions,
as well as those for lepton pairs. For real photons, this exercise has been
carried out recently  \cite{fms}. 
The details of this calculation
will not be discussed here, but they are similar to those in the virtual photon
case, considered now. 
It suffices here to say that the new photon source is dominant, up to $p_{\rm T}
\approx$ 6 GeV/c. Since lepton pair spectra offer the added flexibility of
having two independent variables (invariant mass and three-momentum, for
example), it is imperative to investigate the effect of the new production
mechanism. 

\begin{figure}[!h]
\begin{center}
\includegraphics*[width=8cm,angle=0]{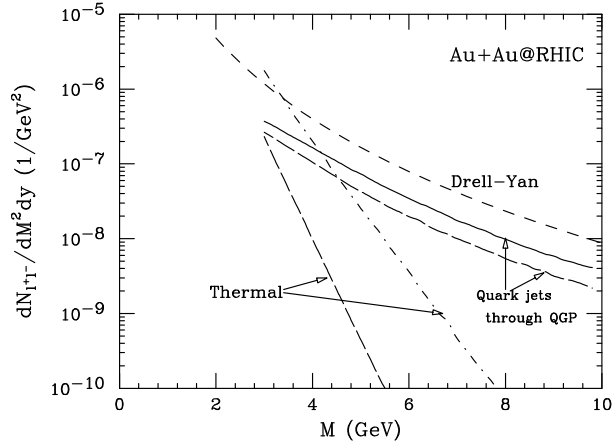}
\caption{Dilepton spectrum from central Pb + Pb collisions at 
$\sqrt{s_{\rm nn}}$ = 200 GeV. The
contributions from Drell-Yan, thermal, and jet-plasma interactions are shown.
More details in the text and references.} 
\label{leptRHIC}
\end{center}
\end{figure}

\begin{figure}[!h]
\begin{center}
\includegraphics*[width=8cm,angle=0]{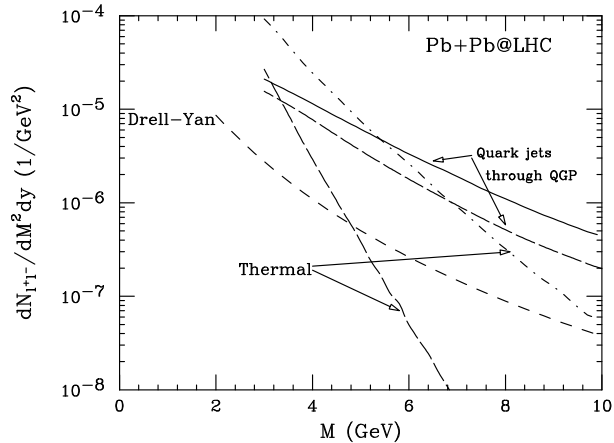}
\caption{Same as Fig. \protect\ref{leptRHIC}, for Pb + Pb collisions at 
$\sqrt{s_{\rm nn}}$ = 5.5 TeV.} 
\label{leptLHC}
\end{center}
\end{figure}

We consider lepton pairs produced by quark-antiquark annihilation, where one
parton is thermal and the other is a hard QCD jet. The obvious background to our
new signal will be that from Drell-Yan processes, which we also compute. We have
estimated the jet and Drell-Yan production in lowest order pQCD, using CTEQ5L
parton distributions and EKS98 shadowing corrections \cite{sgf}. As a first
step, a thermally and chemically equilibrated plasma is assumed to be created at
proper time $\tau_0$. An isentropic expansion will yield the usual correlation
between the measured charged particle multiplicity and initial temperature
\cite{bj}. The details are to be found in \cite{sgf}. One obtains the result
shown in Fig. \ref{leptRHIC}, for symmetric collisions of lead nuclei at RHIC
energies.
The two curves for the thermal and jet-plasma interactions are associated with
$\tau_0$ = 0.15 and 0.5 fm/c, top and bottom respectively. The higher
temperature (shorter formation time) represents a rapid thermalization limit
set by the uncertainty principle \cite{kms}.  We have explored the sensitivity
of our results to this assumption, by looking also at perhaps more conservative
initial conditions. In fact, we see that the new signal is relatively robust
with respect to temperature variations, as it only depends linearly on the
thermal distributions. This is not the case for the purely thermal radiation, a
``classic'' plasma signature. At RHIC, jet-plasma interactions produce more virtual
photons than the annihilation of thermal partons, but less than the Drell-Yan
mechanism. However, the results at LHC energies, Fig. \ref{leptLHC}, show that
the larger initial temperatures lead to an excess of large mass dileptons over
those from conventional sources. Note that the usual contribution from
correlated heavy quark decays is not shown here.

\begin{figure}[!h]
\begin{center}
\includegraphics*[width=8cm,angle=0]{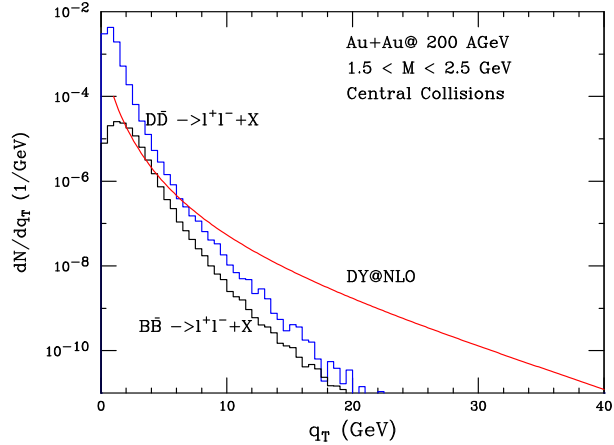}
\caption{Dilepton signal associated with a recoiling jet compared with spectra 
from correlated heavy quark ($c$ and $b$) decays, at RHIC energies.} 
\label{tag1}
\end{center}
\end{figure}

Another aspect of jets and electromagnetic radiation we wish to highlight is the
following. At the parton level, the processes $q g \to q \gamma^*$ and $q
\bar{q} \to g \gamma^*$, are next-to-leading order (in $\alpha_s$), however
those are very effective in producing large momentum lepton pairs. These will be
correlated with a recoiling parton which will traverse and, in principle,
interact with an eventual plasma. The jets will thus have a dilepton tag which,
combined with a jet reconstruction measurement, could yield valuable information about
in-medium energy modification. Photon-tagged jets have been suggested as a probe
for jet-quenching studies \cite{jetq}, but we argue that virtual
photons offer some advantages \cite{sga}. This last statement is illustrated in
Fig. \ref{tag1}, where the lepton pair signal (DY@NLO) is compared to the known
background from correlated heavy quark decays. 
Clearly, there exists a window in
invariant mass and transverse momentum where the electromagnetic signal
from the jet tag outshines that from the background. This assertion holds true at RHIC (Fig.
\ref{tag1}) and, to a somewhat lesser extent, at the LHC \cite{sga}. 

\section{Summary}
Jet-plasma interactions provide a new source of virtual photons that can
compete with the Drell-Yan process and even outshine it, depending on the
energy. In a somewhat more utilitarian vein, dileptons can also be used to tag
jets and thus provide valuable information on their characteristics at formation time. 

\section*{Acknowledgments}
C.G. is supported in part by the Natural Sciences and Engineering Research
Council of Canada, and in part by the Fonds Nature et Technologies of Quebec. 
ORNL is managed by UT-Battelle, LLC, for the U.S. Department of Energy under
contract DE-AC05-00OR22725. R.J.F. is supported by DOE grant DE-FG02-87ER40328.

\end{document}